\documentclass[11pt]{article}
\usepackage{amsmath}
\usepackage{amssymb}
\usepackage{amsthm}
\usepackage{amsfonts}
\usepackage{graphicx}
\usepackage{textcomp}
\usepackage{color}
\usepackage{hyperref}
\numberwithin{equation}{section}

\begin{document}

\begin{titlepage}

\title{New Improved Massive Gravity \\ And\\  Three Dimensional Spacetimes Of\\ Constant Curvature And Constant Torsion}

\author{ Tekin Dereli\footnote{tdereli@ku.edu.tr} \, , Cem Yeti\c{s}mi\c{s}o\u{g}lu\footnote{cyetismisoglu@ku.edu.tr} \\ {\small Department of Physics, Ko\c{c} University, 34450 Sar{\i}yer, \.{I}stanbul, Turkey }}

\date{     }

\maketitle

\begin{abstract}

\noindent We derive the field equations for topologically massive gravity coupled with the most general quadratic curvature terms using the language of exterior differential forms and a first order constrained variational principle. We find variational field equations both in the presence and absence of torsion. We then show that spaces of constant negative curvature 
(i.e. the anti de-Sitter space  $AdS_3$) and constant torsion provide exact solutions. 

\end{abstract}

\vskip 2cm

 Minimal massive gravity. 3D spaces of constant curvature and constant torsion.

\end{titlepage}

\maketitle			% automatic title!
\clearpage 

\section{Introduction}

\noindent  It is often useful to study lower dimensional field theoretic models to gain further insight into fundamental interactions of nature. 
In particular, gravity in (1+2)-dimensions has received a lot of attention as a theoretical tool that highlights the topological aspects of gravitation.
Basic questions such as  whether if gravitational interactions may have a finite range \cite{Boulware-Deser}, or in which sense a quantum gravity might be useful  \cite{leutwyler} led to
insightful answers with this approach. In fact it is well known that Einstein's gravity in (1+2)-dimensions has no dynamics on its own\cite{deser-jackiw-tHooft}.  One may introduce  gravitational degrees of freedom that propagate, either by elevating the gravitational field equations to third order by including in the action a topological Chern-Simons term \cite{deser-jackiw-templeton2, deser-jackiw-templeton1}, or   by coupling other fields such as a dilaton scalar \cite{dereli-tucker1} or a gravitino field \cite{dereli-deser}   
to gravity. Topologically massive gravity (TMG) proved to be interesting since it admits a stationary, circularly symmetric solution 
that is asymptotically $AdS_3$, and behave as if it is a rotating black hole \cite{BTZ1,BTZ2}. 
In many respects, BTZ solution of TMG is the analog of Kerr solution in (1+3)-dimensions. The construction of conserved quantities associated with the BTZ solution \cite{ brown-henneaux, deser-tekin2} and the study  of hidden dualities \cite{dereli-obukhov, carlip et al1} prove to be challenging problems in their own right.   
More recently, unitary extensions of TMG were sought  by the addition of quadratic curvature terms to the action, thus raising the order of the Einstein field equations from three in TMG to four \cite{Deser1, strominger et al, nakasone-oda}. A remarkable extension of TMG, that is called  New Massive Gravity (NMG) in the literature \cite{bergshoeff-hohm-townsend1, bergshoeff-hohm-townsend2}  consists of discarding the Chern-Simons term altogether in the action, so that there are no third derivatives left in the field equations, and replacing it by a particular combination of quadratic curvature invariants that leads to fourth order field equations,  but with unitarity guaranteed at least at the linearized approximation. 
It should be remarked that, all the models of 3D gravity discussed upto this point 
involve (pseudo)-Riemannian space-times. A Minimal Massive Gravity (MMG) was introduced a couple of years  ago for which the variation of the action is done under a non-linear constraint that induces a dynamical space-time torsion \cite{bergshoeff et al4, bergshoeff et al3, baykal}. Several aspects of MMG such as unitarity \cite{tekin, Alishahiha et al}, or its conserved quantities \cite{ altas-tekin2} and exact solutions \cite{arvanitakis, altas-tekin1} have been the subject of very recent studies.   

\noindent Here in what follows,  the field equations for the Einstein-Chern-Simons  gravity coupled with the most general quadratic curvature terms in the action  are derived by a  first order  constrained variational principle. We make  extensive use of  the concise language  of exterior differential forms. Variational field equations both in the presence and absence of torsion are determined. 
%We then discuss space-times of constant negative curvature 
%(i.e.$AdS_3$) and constant torsion as exact background solutions. The notion of constant torsion in 3-dimensions is not new \cite{dereli-vercin2}
%but has been overlooked.
The notion of the torsion of a material continuum has been first introduced by \'{E}.Cartan in 1922\cite{cartan}. This idea found important physical applications, on the one hand, within the context of modified theories of gravity\cite{trautman}, while on the other hand within the context of gauge theories of continuum dislocation and disclination defects\cite{hehl}.
Here we will be discussing 3-dimensional space-times of constant negative curvature 
(i.e.$AdS_3$) and constant torsion\cite{dereli-vercin1,dereli-vercin2} as exact background solutions. The concept of 3-dimensional spaces of constant torsion  was implicit in Cartan's work and it is called Cartan's spiral staircase in a recent review paper\cite{hehl}. 

\bigskip
%\newpage
%%%%%%%%%%%%%%%%%%%%%%%%%%%%%%%%%%%%%%%%%%%%%%%%%%%%%%%%%%%%%

\noindent Notation and Conventions:

\medskip

\noindent Throughout our work, we will be using the language of exterior differential forms.  The metric tensor of space-time, given by  $g= \eta_{ab} e^a \otimes e^b$  where $\eta_{ab}= g(X_a,X_b) = diag(-,+,+)$ is written in terms of an orthonormal basis of frame vectors  $\{X_a\}$ that are dual to the co-frame 1-forms $\{e^a\}$ so that  $e^a(X_b)=\delta^a_b$.    $\iota_a = \iota_{X_a}$  stands for the interior product operators with respect to frame vectors $X_a$.  $*:\Lambda^p(M) \to \Lambda^{3-p} (M)$ denotes the Hodge duality operator acting on $p$-forms.  The space-time orientation is fixed with the choice of the volume 3-form $*1=e^0 \wedge e^1 \wedge e^2$. For convenience, the following abbreviation for the exterior products $e^{ab\dots}=e^a \wedge e^b \wedge \dots $ is going to be used.  
A linear connection on space-time will be specified by a set of  connection 1-forms $\{\omega^{a}_{\;\;b} \}$ .  We will work with a connection that is compatible with the metric but need not be torsion-free.   Then the index raising and lowering operations commute with the covariant exterior derivativation and we have $ D(\omega)\eta_{ab} = \omega_{ab} + \omega_{ba} =0$. 
We specify the torsion 2-forms   $T^a$ of space-time through the  first set of Cartan structure equations 
\begin{equation}
d e^a + \omega^a_{\ b} \wedge e^b = T^a ,
\end{equation}
while  the curvature 2-forms $R^a_{\ b}(\omega)$ through the second set of Cartan structure equations 
\begin{equation}
d\omega^a_{\ b} + \omega^a_{\ c} \wedge \omega^c_{\ b} = R^a_{\ b}(\omega).
\end{equation}
The following Bianchi identities are obtained as  integrability conditions of  the above Cartan structure equations: 
\begin{equation}
D(\omega)T^a= R^a_{\ b}(\omega) \wedge e^b, \quad 
D(\omega) R^a_{\ b}(\omega) = 0.
\end{equation}
It is convenient to define contortion 1-forms $K^a_{\ b}$  as the difference between our non-Riemannian connection 1-forms   and the Riemannian (Levi-Civita) connection 1-forms  
$\{{\hat{\omega}}^a_{\ b}\}$ that satisfy the structure equations
\begin{equation}
d e^a + {\hat{\omega}^a_{\ b}} \wedge e^b = 0 .
\end{equation}
Thus we have 
\begin{equation}
K^a_{\ b} = \omega^a_{\ b} - {\hat{\omega}^a_{\ b}}
\end{equation}
which are in one to one correspondence with the  torsion 2-forms $T^a$ through the structure equations
\begin{equation}
K^a_{\ b} \wedge e^b = T^a ,
\end{equation}
or conversely
\begin{equation}
2 K_{ab} = \iota_{a} T_{b} - \iota_{b} T_{a}  - e^{c} \iota_{ab}T_{c}. 
\end{equation}
It is not difficult to find a relation between the non-Riemannian curvature 2-forms $R^a_{\ b}(\omega)$ and the Riemannian curvature 2-forms $R^a_{\ b}(\hat{\omega})$ of the Levi-Civita connection as 
\begin{equation}
R^a_{\ b}(\omega)= R^a_{\ b}(\hat{\omega}) + D(\hat{\omega}) K^a_{\ b}+K^a_{\ c} \wedge K^c_{\ b}
\end{equation} 
where $D(\hat{\omega})$ denotes the covariant exterior derivative with respect to  the Levi-Civita connection. 
The Ricci 1-forms $Ric_a = {\cal{R}}_{ab}e^b $ and the curvature scalar ${\cal{R}}$ are obtained by contractions of the curvature 2-forms as follows:
\begin{equation}
Ric_a = \iota^b R_{ba} ,\quad
{\cal{R}} = \iota^a Ric_a = \iota^{ab} R_{ba}.
\end{equation}
Moreover, the Einstein 2-forms of our non-Riemannian connection are defined by 
\begin{equation}
G_a (\omega)= G_{ab} *e^b = *Ric_a - \frac{1}{2} {\cal{R}} *e_a  = -\frac{1}{2} R^{bc}(\omega) * e_{abc} .
\end{equation}
We note that in 3-dimensions, the curvature 2-forms are in one to one correspondence with the Einstein 2-forms   since
\begin{equation}
\epsilon_{abc}  G^{c}(\omega) =  R_{ab}(\omega) 
\end{equation}
where $\epsilon_{abc}$ denotes the completely anti-symmetric Levi-Civita  symbol in three dimensions with $\epsilon_{012}=1$. 
We may therefore give the curvature 2-forms $R_{ab}$ in 3-dimensions in terms of the Ricci 1-forms $Ric_a$ and the curvature scalar ${\cal{R}}$:
\begin{equation}
R_{ab} = \epsilon^c_{\ ab} * Ric_c +\frac{1}{2}{\cal{R}} e_a \wedge e_b .
\end{equation}
As a consequence, quadratic curvature invariants in 3-dimensions are related to each other through the identity
\begin{equation}
R^{ab} \wedge* R_{ab} - 2Ric^a \wedge * Ric_a +\frac{1}{2} {\cal{R}}^2 *1 = 0 \label{identity1},
\end{equation}
that is, any one of the quadratic curvature invariants in 3-dimensions can be expressed in terms of the other two. 
%%%%%%%%%%%%%%%%%%%%%%%%%%%%%%%%%%%%%%%%%%%%%%%%%%%%%%%%%%%%%%%%%%%%%%%%
\section{Action}

\noindent  The field equations of our model will be determined by the constrained variations of an action integral
\begin{equation}
I[e^a,\omega^a_{\ b}, \lambda_a]= \int_M \mathcal{L}
\end{equation}
where $M$ is a compact region on some chart on a (1+2)-dimensional Riemann-Cartan manifold. The independent variables on which the action depends are the  co-frame 1-forms $\{e^a\}$, connection 1-forms $\{ \omega^a_{\ b} \}$, and Lagrange multiplier 1-forms $\{\lambda_a\}$. We consider a Lagrangian density 3-form 
\begin{equation}
\mathcal{L}={\mathcal{L}}_{TMG} + {\mathcal{L}}_{2} +{\mathcal{L}}_{C}
\end{equation}
where
\begin{equation}
{\mathcal{L}}_{TMG}=\frac{1}{\mu}(\omega^a_{\ b} \wedge d\omega^b_{\ a} + \frac{2}{3} \omega^a_{\ b} \wedge \omega^b_{\ c} \wedge \omega^c_{\ a}) + \frac{1}{2K} R^{ab} \wedge *e_{ab} + \Lambda *1 
\end{equation}
is the Lagrangian density of the topologically massive gravity (TMG);
\begin{equation}
{\cal{L}}_{2} = \alpha R^{ab} \wedge *R_{ab}+ \beta Ric^a \wedge *Ric_a + \gamma {\cal{R}}^2 *1 
\end{equation} 
is a generic quadratic curvature term with three coupling constants $\alpha$, $\beta$, and $\gamma$. It should be remarked that, there are alternative ways of 
specifying a generic quadratic curvature invariant in three dimensions. Due to the identity (\ref{identity1}), either one of the terms $ R^{ab} \wedge *R_{ab}$, $Ric^a \wedge *Ric_a$ or $ {\cal{R}}^2 *1 $ may be made redundant  in favor of others. Therefore, still keeping the coupling constants $\alpha$, $\beta$ and $\gamma$, we may consider without loss of generality, any one of the following combinations:
\begin{align}
{\cal{L}}_{2} &= (2\alpha + \beta) Ric^a \wedge *Ric_a + (\gamma - \frac{\alpha}{2})  {\cal{R}}^2 *1  \nonumber \\
{\cal{L}}_{2} &= (\alpha + \frac{\beta}{2}) R^{ab} \wedge *R_{ab}+ (\gamma + \frac{\beta}{4})  {\cal{R}}^2 *1  \nonumber \\
{\cal{L}}_{2} &= (\alpha-2\gamma)  R^{ab} \wedge *R_{ab}+ (\beta+4\gamma)  Ric^a \wedge *Ric_a   .
\end{align} 
For technical ease. we prefer to work with the second alternative.  Finally,
\begin{equation}
{\mathcal{L}}_{C} = T^a \wedge \lambda_a + \frac{\nu}{2} \lambda_a \wedge \lambda_b \wedge *e^{ab}  
  \end{equation}
 is the constraint Lagrangian density 3-form, which in case $\nu=0$ imposes the constraint that the connection is the torsion-free Levi-Civita connection. On the other hand if $ \nu \neq 0$, the torsion 2-forms 
 would be related with the Lagrange multiplier 1-forms in a non-trivial way. 
 All the previously studied models such as TMG, NMG or MMG are  covered as sub-cases with the  choice (2.2) of the action .

%%%%%%%%%%%%%%%%%%%%%%%%%%%%%%%%%%%%%%%%%%%%%%%%%%%%%%%%%%%%%%%%%%%%%%
\section{Variational Field Equations}

\noindent We evaluate the variational derivative  of the total Lagrangian and find  (upto a closed form)
\begin{align}
{\dot{\cal{L}}} &= {\dot{e}}^a \wedge \bigg\{ \frac{1}{2K} R^{bc} \wedge * e_{abc} +\Lambda *e_a  +(\alpha+\frac{\beta}{2})  \bigg ( \iota_a R^{bc} \wedge *R_{bc} -R^{bc} \wedge \iota_a *R_{bc} \bigg )
\nonumber \\ 
&\qquad  + (\gamma + \frac{\beta}{4}) \bigg (2 {\cal{R}} R^{bc} \wedge * e_{abc} - {\cal{R}}^2 * e_a \bigg ) \nonumber \\  
& \qquad + D(\omega)\lambda_a + \frac{\nu}{2} \lambda^b \wedge \lambda^c \wedge *e_{abc} \bigg\} \nonumber  \\
& \qquad +{\dot{\omega}}^{ab} \wedge \bigg\{ \frac{2}{\mu} R_{ba} + \frac{1}{2K} T^c \wedge * e_{abc} + (2\alpha+\beta)D(\omega)*R_{ab}  \nonumber \\
&\qquad + (2\gamma+\frac{\beta}{2})  D(\omega)({\cal{R}}  *e_{ab}) +\frac{1}{2} (e_b \wedge \lambda_a - e_a \wedge \lambda_b ) \bigg\} \nonumber \\
& \qquad +{\dot{\lambda}}_a \wedge \bigg \{T^a  + \nu \lambda_{b} \wedge *e^{ab}   \bigg \} .
\end{align}
Here a dot over a field variable denotes the variation of the corresponding field. We first impose the constraint
\begin{equation}
T^a = -\nu \lambda_{b} \wedge *e^{ab} \iff  K_{ab} = \nu \epsilon_{abc} \lambda^{c}.
\end{equation} 
Then we go to connection variation equations which now read
\begin{equation}
e_a \wedge \lambda_b - e_b \wedge \lambda_a  = Q^{-1} \Sigma_{ab} \label{lambda1}
\end{equation}
where we set
\begin{equation}
Q = \frac{1}{2} - \frac{\nu}{2K} - \nu (2\gamma +\frac{\beta}{2}) {\cal{R}}, \label{qu}
\end{equation}
and 
\begin{equation}
\Sigma_{ab} = -\frac{2}{\mu}R_{ab} +(2\alpha +\beta)D(\omega)*R_{ab} + (2\gamma+\frac{\beta}{2}) d{\cal{R}} \wedge *e_{ab}.
\end{equation}
 We  solve (\ref{lambda1}) algebraically for the Lagrange multiplier 1-forms and determine
\begin{equation}
\lambda_a = Q^{-1} \bigg ( -\frac{2}{\mu} Y_a  + (2 \alpha+\beta) W_a + (2\gamma+\frac{\beta}{2}) ( \iota_{a}*d{\cal{R}})  \bigg ) \label{lambda2}
\end{equation}
where we introduced further abbreviations
\begin{equation}
Y_a = Ric_a - \frac{1}{4}e_a {\cal{R}},  \qquad
W_a = \iota^{b}(D(\omega)*R_{ba})  - \frac{1}{4}e_a (\iota^b \iota^c D(\omega)*R_{cb}). 
\end{equation}
Finally we substitute (\ref{lambda2}) into the  co-frame variation equations and arrive at the Einstein field equations given as follows:
\begin{align}
&\bigg (\frac{1}{2K} +(2\gamma+\frac{\beta}{2}){\cal{R }} \bigg) R^{bc} \wedge * e_{abc} + \bigg ( \Lambda -(\gamma+\frac{\beta}{4}){\cal{R}}^2 \bigg)  *e_a \nonumber \\
&+ (\alpha +\frac{\beta}{2}) \bigg ( \iota_a R^{bc} \wedge *R_{bc} -R^{bc} \wedge \iota_a *R_{bc} \bigg )  \nonumber\\
&D\lambda_a + \frac{\nu}{2} \lambda^b \wedge \lambda^c *e_{abc} = 0 .
\end{align}
We note that these equations include among other terms, the Cotton-Schouten 2-forms
\begin{equation}
C_a \equiv  D(\omega)Y_a = D(\omega)(Ric_a - \frac{1}{4}{\cal{R}} e_a)
\end{equation}
that involve third derivatives of the metric components and the 2-forms 
\begin{equation}
D_a \equiv D(\omega)W_a = D(\omega)( \iota^{b}(D(\omega)*R_{ba})  - \frac{1}{4}e_a  \iota^b \iota^c (D(\omega)*R_{cb}) )
\end{equation}
that involve fourth  derivatives of the metric components.
We also note that Einstein field equations in the case of zero-torsion ($\nu=0$) are given by
\begin{align}
&\bigg (\frac{1}{2K} +(2\gamma+\frac{\beta}{2}){\hat{\cal{R }}} \bigg) {\hat{R}}^{bc} \wedge * e_{abc} + \bigg ( \Lambda -(\gamma+\frac{\beta}{4}){\hat{\cal{R}}}^2 \bigg)  *e_a \nonumber \\
&+ (\alpha +\frac{\beta}{2}) \bigg ( \iota_a {\hat{R}}^{bc} \wedge *{\hat{R}}_{bc} -{\hat{R}}^{bc} \wedge \iota_a *{\hat{R}}_{bc} \bigg )  \nonumber\\
& - \frac{4}{\mu}{\hat{C}}_a  + (4\alpha+2\beta){\hat{D}}_a + (4\gamma+\beta) D({\hat{\omega}})(\iota_{a}*d{\hat{\cal{R}}}) = 0 .
\end{align}
Field equations (3.11) go down consistently to the Topologically Massive Gravity (TMG) field equations if the quadratic curvature terms are absent, i.e.  if we set $\alpha =\beta=\gamma=0$ above. 
 Finally we re-write Einstein field equations in two special cases of recent interest: 
 
\noindent New Massive Gravity (NMG):   $\frac{1}{\mu} \rightarrow 0$, $\Lambda=0$, $\alpha=0, \beta=1,  \gamma= -\frac{3}{8},  \nu=0$. 
 \begin{align}
&\bigg (\frac{1}{2K} -\frac{1}{4} {\hat{\cal{R }}} \bigg) {\hat{R}}^{bc} \wedge * e_{abc} +  \frac{1}{8}{\hat{\cal{R}}}^2   *e_a \nonumber \\
& + \frac{1}{2} \bigg ( \iota_a {\hat{R}}^{bc} \wedge *{\hat{R}}_{bc} -{\hat{R}}^{bc} \wedge \iota_a *{\hat{R}}_{bc} \bigg ) 
 +2 {\hat{D}}_a - \frac{1}{2} D({\hat{\omega}}) (\iota_{a}*d{\hat{\cal{R}}}) = 0 .
\end{align}
 
 \noindent Minimal Massive Gravity (MMG): $\alpha =\beta=\gamma=0,  \nu \neq 0$.
\begin{align}
& -\frac{1}{K} G_{a} +  \Lambda *e_a  - \frac{4K}{\mu(K-\nu)} D(\omega)Y_{a} + \frac{8K^2\nu}{\mu^2(K-\nu)^2} Y^b \wedge Y^c \wedge *e_{abc} = 0,\nonumber \\
& K_{ab} = -\frac{4K\nu}{\mu (K-\nu)} \epsilon_{abc} Y^{c} .
 \end{align}

%%%%%%%%%%%%%%%%%%%%%%%%%%%%%%%%%%%%%%%%%%%%%%%%%%%%%%%%%%%%%%%%%%%%%%

\section{Background Solutions with $AdS_3$}

\noindent In order to proceed any further in the study of a 3D quantized theory of  gravity based on our model,  its background solutions should be found.  
Towards that end, here we  consider three dimensional non-Riemannian space-times of constant curvature and constant torsion\cite{dereli-vercin1,dereli-vercin2}. We also conveniently work with coordinate independent methods\cite{dereli-tucker1}. That is to say, we  evaluate curvatures and
their derivatives without differentiating  any functions. The relevant differential geometric techniques are briefly explained in an appendix.
Our starting point will be the structure equations satisfied by an orthonormal  set of left-invariant basis 1-forms $\{e^a\}$ on $AdS_3$:
\begin{equation}
de^a = -  \frac{1}{\rho}\epsilon^{a}_{\;\;\;bc} e^b \wedge e^c .
\end{equation} 
Thus we determine the Levi-Civita connection 1-forms
\begin{equation}
{\hat{\omega}}^{a}_{\;\;b} = - \frac{1}{\rho} \epsilon^{a}_{\;\;bc} e^c,
\end{equation}
and the corresponding curvature 2-forms
\begin{equation}
 {\hat{R}}^{a}_{\;\;b} = - \frac{1}{\rho^2} e^a \wedge e_{b}.
 \end{equation} 
Now, we set the torsion 2-forms to be
\begin{equation}
T^a = \frac{2}{\sigma} *e^a, \quad \sigma^2 \neq \rho^2  \iff K^{a}_{\;\;b} = -\frac{1}{\sigma} \epsilon^{a}_{\;\;bc} e^c.
\end{equation}
Then the full curvature 2-forms turn out to be
\begin{equation}
 R^{a}_{\;\;b} = \big ( \frac{\rho^2 - \sigma^2}{\rho^2 \sigma^2} \big )  e^a \wedge e_{b}.
 \end{equation} 
Their contractions give
\begin{equation}
Ric_a = 2\big (\frac{\rho^2 - \sigma^2}{\rho^2 \sigma^2}\big ) e_a  , \quad {\cal{R}} = 6 \big (\frac{\rho^2 - \sigma^2}{\rho^2 \sigma^2}\big ).
\end{equation} 
Substituting these in (\ref{qu}), we find
\begin{equation}
Q= \frac{1}{2} -\frac{\nu}{2K} -6\nu (2\gamma+\frac{\beta}{2}) \big (\frac{\rho^2-\sigma^2}{\rho^2 \sigma^2}\big),
\end{equation}
and in (3.6), we find
\begin{equation}
\lambda_a = -Q^{-1} \bigg (  \frac{1}{\mu}  + \frac{2\alpha+\beta}{\sigma}  \bigg )  \big ( \frac{\rho^2-\sigma^2}{\rho^2 \sigma^2} \big )  e_a.  
\end{equation}
We must first check (3.2) for consistency:
\begin{equation}
\frac{1}{\nu \sigma} = Q^{-1} \bigg (  \frac{1}{\mu} + \frac{2\alpha+\beta}{\sigma} \bigg ) \big ( \frac{\rho^2-\sigma^2}{\rho^2 \sigma^2} \big ).
\end{equation}
Substituting for $Q$ from (4.7), we get an algebraic consistency equation as follows:
\begin{equation}
2 \bigg (  \frac{\sigma}{\mu} + 2\alpha+4\beta +12 \gamma \bigg ) \big ( \frac{\rho^2-\sigma^2}{\rho^2 \sigma^2} \big ) = \frac{K-\nu}{K \nu}.
\end{equation}
Next we go over to the Einstein field equations (3.8) with
 \begin{equation}
 \lambda_a = -\frac{1}{\nu \sigma}e_a 
 \end{equation}  
 and organise terms to arrive at
\begin{equation}
(2\alpha -2\beta -12 \gamma) \big ( \frac{\rho^2-\sigma^2}{\rho^2 \sigma^2} \big )^2
+ \frac{1}{K} \big ( \frac{\rho^2-\sigma^2}{\rho^2 \sigma^2} \big ) +(\Lambda - \frac{1}{\nu \sigma^2 }) = 0.  
\end{equation}
Thus we obtain two algebraic equations (4.10) and (4.12) that are  to be solved simultaneously for $\rho$ and $\sigma$ in terms of the coupling parameters
$K, \Lambda, \alpha, \beta,\gamma, \mu$ and $\nu$.  
%%%%%%%%%%%%%%
In order to make further headway, we introduce new variables
\begin{equation}
\xi = \frac{1}{\rho}, \quad \eta = \frac{1}{\sigma}, \quad a = \frac{\mu}{4} \big (  \frac{1}{\nu} - \frac{1}{K}\big )
\end{equation}
and  write down the following simultaneous  algebraic equations:
\begin{equation}
\frac{1}{(\xi + \eta)} + \frac{1}{(\xi - \eta)} = \frac{1}{a^2} + \frac{(\alpha+2\beta+6\gamma)}{a} (\xi + \eta) +    \frac{(\alpha+2\beta+6\gamma)}{a} (\xi - \eta),
\end{equation}
\begin{eqnarray}
\big ( \frac{\xi+\eta}{\xi - \eta} \big ) +\big ( \frac{\xi-\eta}{\xi +\eta} \big ) = \big ( \frac{4\nu}{K} -1 \big ) &+& 4\nu (2\alpha -2\beta - 12 \gamma)  (\xi-\eta)(\xi +\eta) \nonumber \\ 
& &+ \frac{4\nu \Lambda}{(\xi-\eta)(\xi +\eta)} .  
\end{eqnarray}
Let us now make a further change of variables
\begin{equation}
t = (\xi-\eta)+(\xi +\eta), \quad s = (\xi-\eta)(\xi +\eta),
\end{equation}  
in terms of which we have
\begin{eqnarray}
at &=& \frac{1}{a}s + (\alpha+2\beta+6\gamma) t s, \\ 
t^2 &=& \big ( \frac{4\nu}{K}+1\big ) s + 4\nu ( 2\alpha-2\beta -12\gamma) s^2 + 4\nu \Lambda.  
\end{eqnarray}
One can now solve the first equation for $t$ in terms of $s$ and substitute it in the second equation.
Thus a quartic  equation for $s$ is reached whose solutions can be obtained in a  standard way. Then working through the equations backwards, 
solutions for $\rho$ and $\sigma$ may be written explicity. As they stand, they are not very instructive. However,  
to be concrete, we concentrate  on a simpler case and  discuss the parameter ranges for the existence of solutions for MMG with $\alpha=\beta=\gamma=0$, 
%We rename  our variables as
%\begin{equation}
%\frac{1}{\sigma} \equiv \xi, \quad \frac{1}{\rho} \equiv \eta,
%\end{equation}
%and let
%\begin{equation}
%a  \equiv \frac{\mu}{4} ( \frac{1}{\nu} - \frac{1}{K} ). 
%\end{equation}
in which  case our coupled system of algebraic  equations reduce to the equations of two  conic sections in the $(\xi\eta)$-plane, given by
\begin{equation}
\frac{(\xi-a)^2}{a^2} -\frac{\eta^2}{a^2} = 1 
\end{equation}
and 
\begin{equation}
 K (\frac{1}{\nu} - \frac{1}{K}) \frac{\xi^2}{K \Lambda} +\frac{\eta^2}{K \Lambda } = 1.
 \end{equation}
 We take $K > 0$ without loss of generality at this point, since our model is not yet coupled to matter.
 We also point out that solutions come in pairs with values $\eta \leftrightarrow -\eta$, as a change in sign of $\eta$ means going from one orientation of the  co-basis to the other, or vice versa . 
 In what follows, we restrict attention to the cases $ 0 <\eta$, but extension to cases $\eta <0$ is easy. 
 Then we classify possible pairs $(\xi, \eta)$ in accordance with the following ranges of our free parameters:
 \begin{itemize}
  \item For $\Lambda < 0$ and $-\infty < \mu < \infty$;   no solution exists with  $0 \leq \nu \leq K$. 
  % \item $\Lambda < 0, \quad \mu <0, \quad 0 \leq \nu \leq K .$\\ There is no solution. 
   \item $\Lambda < 0, \quad \mu >0, \quad \nu \leq 0 \; $or $\; K \leq \nu.$\\ There is a solution for $0 \leq \xi \leq \sqrt{|\frac{(K-\nu)}{\nu} K\Lambda | }$  and $0 < \eta < \infty$.\\
   A second  one may exist for  $-\sqrt{|\frac{(K-\nu)}{\nu}K\Lambda |} \leq \xi \leq - \mu |\frac{(K-\nu)}{2K\nu}|$, depending on the magnitude of $\mu$.   
  \item $\Lambda < 0, \quad \mu <0, \quad  \nu \leq  0 \;$ or $\; K \leq \nu. $\\ There is a solution for $ \xi \leq -\sqrt{ |\frac{(K-\nu)}{\nu} K\Lambda | }$  
  and $ 0 < \eta < \infty$.\\  A second one may exist for  $\sqrt{|\frac{(K-\nu)}{\nu}K\Lambda |} \leq \xi \leq \mu \frac{(K-\nu)}{2K\nu}$, depending on the magnitude of $\mu$.   
  \item $\Lambda >0, \quad \mu >0, \quad 0 \leq \nu \leq K .$\\ Solutions exist for $ 0 \leq \eta \leq \sqrt{K\Lambda}$. Then there is a solution for $-\sqrt{\frac{K-\nu}{\nu}K\Lambda} \leq \xi \leq 0.$ A second one may exist , depending on the magnitude of $\mu$, for   $\sqrt{\frac{K-\nu}{\nu}K\Lambda} \leq \xi \leq \mu \frac{(K-\nu)}{2K\nu}.$
  \item $\Lambda > 0, \quad \mu <0, \quad \nu \leq 0 \;$ or $\; K \leq \nu.$\\  Two solutions  exist for $\sqrt{K\Lambda} \leq \eta$ and with either $\xi \leq 0$
  or $\mu\frac{(K-\nu)}{2K\nu} \leq \xi.$
  \item $\Lambda  >0, \quad \mu >0, \quad  \nu \leq 0  \;$ or $\; K  \leq \nu .$\\  Two solutions exist for $\sqrt{K\Lambda} \leq \eta$ and with either  $\xi \leq  -\mu\frac{(K-\nu)}{2K\nu}$ or $0 \leq \xi.$
  \item $\Lambda >0, \quad \mu  < 0, \quad 0 \leq \nu \leq K. $\\ Solutions exist for  $\sqrt{K\Lambda} \leq \eta$.  One  solution has $0 \leq \xi \leq \sqrt{\frac{\nu}{K-\nu} K\Lambda}.$ A second one may exist, depending on the magnitude of $\mu$, if  $ -\sqrt{\frac{\nu}{K-\nu} K\Lambda} \leq \xi \leq -\sqrt{|\mu|\frac{(K-\nu)}{\nu}}.$ 
 \end{itemize}
 
\bigskip
%%%%%%%%%%%%%%%%%%%%%%%%%%%%%%%%%%%%%%%%%%%%%%%%%%%%%%%%%%%%%%%%%%%%%%%
%\newpage

\section{Concluding Remarks}

\noindent  In this paper, we considered the extension of  Einstein-Chern-Simons  gravity (TMG)  with the most general quadratic curvature terms in the action and derived the corresponding field equations by a  first order  constrained variational principle. We made  extensive use of  the concise language  of exterior differential forms. Variational field equations 
were determined  both in the absence and presence of a dynamical space-time torsion. 
It should be emphasised  that our discussion  based on the choice (2.2) of the action encompasses all currently studied models such as NMG or MMG as particular sub-cases. 
 In order to specify the ground state of our model we then  considered Riemann-Cartan space-times of constant negative curvature 
(i.e.$AdS_3$) and constant torsion as exact background solutions. 
\newline Finally we wish to add the following comments:

1. In recent literature, the generic quadratic curvature term in the action that is commonly used  is given by the first alternative in Eqn.(2.5). Here we use the 
second alternative for technical ease and were able to present the final field equations (3.8) in a compact and geometrically transparent way. 

2. The notion of 3-dimensional Riemann-Cartan spaces with constant curvature and constant torsion  is not new \cite{dereli-vercin1,dereli-vercin2}, but not used in  this context before. 
It provides a novel pathway for the construction of physically relevant 3D-gravity configurations.  

3. The coupled algebraic equations (4.14) and (4.15)  describe a cubic and a quartic curve, respectively, in the $(\xi \eta)$-plane. 
We showed the existence of  intersection points in the  special case of MMG ($\alpha=\beta=\gamma=0$) where  both curves reduce to conic sections.

\bigskip

\section{Acknowledgement}

\noindent C.Y. thanks Ko\c{c} University for a Graduate Student Scholarship.

%\newpage

\bigskip

\section{Appendix}

\noindent $AdS_3$ can be realized as an embedded hypersurface in a four dimensional flat space equipped  with an indefinite metric $g= -dU^2-dV^2+dX^2+dY^2$, written  
in Cartesian coordinates $\{\xi^A ; A=1,2,3,4 \} : (U,V,X,Y)$. In the same coordinate system,  the embedding equation will be given by 
\begin{equation}
-U^2-V^2+X^2+Y^2=-1.
\end{equation}
Furthermore we know that $AdS_3$ is an homogeneous space with  $AdS_3=SO(2,2)/SO(2,1)$. 
In order to  verify that the Lie algebra  $so(2,2)$ of isometries  is a direct product of two copies of $so(2,1)$, we consider, in our Cartesian system $\xi_A:(-U,-V,X,Y)$, the Killing vector fields $J_{AB}$ that are given explicitly by,
\begin{equation}
J_{AB}=\xi_A\frac{\partial}{\partial \xi^B} -\xi_B\frac{\partial}{\partial \xi^A}
\end{equation}
and satisfy the commutation relations:
\begin{eqnarray}
[J_{AB},J_{ BC}]= \begin{cases} -J_{AC},  & \mbox{for}\  \ B \in \{1,2\}\ \mbox{and}\ A\neq B \neq C, \\
J_{AC},  & \mbox{for}\  \ B \in \{3,4\}\ \mbox{and}\ A\neq B \neq C.
\end{cases}
\end{eqnarray}
It is straightforward to divide these Killing vector fields into two conjugacy classes by defining the left-invariant vector fields 
\begin{equation}
X_0 = -J_{UV}-J_{XY}, \quad X_1 = J_{XU}+J_{YV}, \quad  X_2 = J_{YU}-J_{XV}, 
\end{equation}
and the right-invariant vector fields
\begin{equation}
 Y_0 = -J_{UV}+J_{XY}, \quad
  Y_1 = J_{XU}-J_{YV}, \quad
 Y_2 = -J_{YU}-J_{XV}
\end{equation}
Both the left-invariant vector fields $\{ X_a: a=0,1,2 \}$ and the right-invariant vector fields $\{ Y_a: a=0,1,2 \}$ satisfy the same commutation relations
\begin{align}
[ X_0, X_1 ] &=  2X_2,\ \ \ [X_1, X_2] = -2 X_0, \ \  \ [ X_0, X_2]= -2 X_1,	\\
[ Y_0, Y_1 ] &= 2 Y_2,\ \ \ \ [Y_1, Y_2] = -2 Y_0, \ \  \ \ \ [ Y_0, Y_2]= -2 Y_1,	
\end{align}
and they commute with each other, i.e.
\begin{equation}
[X_a,Y_b] =0, \quad a,b=0,1,2.
\end{equation}
At this point, we choose a local coordinate chart $x^{\mu}:(t,\chi, \theta)$ for $AdS_3$ such that 
\begin{equation} 
U= \cos t ,   V= \sin t \cosh\chi ,  \ X= \sin t \sinh \chi \cos \theta,  Y= \sin t \sinh \chi \sin\theta.
\end{equation}
Then we compute  the following explicit expressions for $\{ X_a \}$:
\begin{eqnarray}
X_0& =&\cosh \chi \partial_t - \cot t \sinh \chi \partial_\chi - \partial_\theta, \nonumber \\
 X_1 &=& - \sinh \chi \cos \theta \partial_t + (\cot t \cosh \chi \cos \theta + \sin\theta)\partial_\chi  \nonumber \\ & & + (  coth \chi \cos \theta-\cot t cosech \chi \sin\theta)\partial_\theta, \nonumber \\
 X_2 &=& - \sinh \chi \sin\theta \partial_t + (\cot t \cosh \chi \sin\theta - \cos \theta)\partial_\chi  \nonumber \\ & & + ( coth \chi \sin\theta+\cot t cosech \chi \cos \theta )\partial_\theta, \nonumber
\end{eqnarray}
and for $\{Y_a \}$:
\begin{eqnarray}
Y_0& =&\cosh \chi \partial_t - \cot t \sinh \chi \partial_\chi + \partial_\theta, \nonumber \\
 Y_1 &=& - \sinh \chi \cos \theta \partial_t + (\cot t \cosh \chi \cos \theta - \sin\theta)\partial_\chi  \nonumber \\ & & + ( - coth \chi \cos \theta-\cot t cosech \chi \sin\theta)\partial_\theta, \nonumber \\
 Y_2 &=&  \sinh \chi \sin\theta \partial_t + (-\cot t \cosh \chi \sin\theta - \cos \theta)\partial_\chi  \nonumber \\ & & + ( coth \chi \sin\theta-\cot t cosech \chi \cos \theta )\partial_\theta. \nonumber
\end{eqnarray}
Finally, exploiting  the dualities $e^b(X_a) = \delta^b_a$ and $\tilde{e}^b(Y_a) = \delta^b_a$, we determine in a unique way  the following set of left-invariant co-frame 1-forms:
\begin{eqnarray}
e^0 & =& \cosh \chi dt +  \cos t \sin t \sinh \chi d\chi + \sin^2 t \sinh^2 \chi d\theta,\\
 e^1 &=& \sinh \chi \cos \theta dt + ( \cos t \sin t \cosh \chi \cos \theta + \sin^2 t \sin\theta)d\chi \nonumber\\
 &+& \sin^2 t  \sinh \chi( \cosh \chi \cos \theta - \cot t  \sin\theta)d\theta,\\
 e^2 &=& \sinh \chi \sin\theta dt + ( \cos t \sin t \cosh \chi \sin\theta - \sin^2 t \cos \theta)d\chi \nonumber\\
 &+& \sin^2 t  \sinh \chi( \cosh \chi \sin\theta + \cot t  \cos \theta)d\theta,
\end{eqnarray}
and the right-invariant co-frame 1-forms:
\begin{eqnarray}
\tilde{e}^0 & =& \cosh \chi dt +  \cos t \sin t \sinh \chi d\chi - \sin^2 t \sinh^2 \chi d\theta,\\
\tilde{e}^1 &=& \sinh \chi \cos \theta dt + ( \cos t \sin t \cosh \chi \cos \theta - \sin^2 t \sin\theta)d\chi \nonumber\\
 &-& \sin^2 t  \sinh \chi( \cosh \chi \cos \theta + \cot t  \sin\theta)d\theta,\\
\tilde{e}^2 &=&- \sinh \chi \sin\theta dt - ( \cos t \sin t \cosh \chi \sin\theta + \sin^2 t \cos \theta)d\chi \nonumber\\
 &+& \sin^2 t  \sinh \chi( \cosh \chi \sin\theta - \cot t  \cos \theta)d\theta,
\end{eqnarray}
It is now straightforward to verify i) that these basis 1-forms satisfy the first Cartan structure equations
\begin{equation}
de^a =  -\epsilon^{a}_{\;\; bc} e^b \wedge e^c \ \ , \ \ d\tilde{e}^a =  -\epsilon^{a}_{\;\; bc} \tilde{e}^b \wedge \tilde{e}^c ,
\end{equation}
and that ii) in our local coordinate chart the metric tensor becomes
\begin{align}
g_{AdS_3} &= -e^0 \otimes e^0 +e^1 \otimes e^1 +e^2 \otimes e^2 \nonumber\\
&=  -\tilde{e}^0 \otimes \tilde{e}^0 +\tilde{e}^1 \otimes \tilde{e}^1 +\tilde{e}^2 \otimes \tilde{e}^2 \nonumber\\
&= -dt^2 + \sin^2t( d\chi^2 + \sinh^2 \chi d\theta^2).
\end{align}

\noindent As a further remark, suppose we let  the right-invariant vector fields change sign i.e.  $Y_a \mapsto W_a= -Y_a \ , \ a=0, 1, 2$. Note that the volume form also switches sign.
Then the  vector fields $\{ W_a \}$  commute with the left-invariant vector fields $\{ X_a \}$, but their structure constants get modified to $-2\epsilon_{abc}$. These new vector fields are explicitly written as :
\begin{eqnarray}
W_0& =&-\cosh \chi \partial_t + \cot t \sinh \chi \partial_\chi - \partial_\theta, \nonumber \\
 W_1 &=&  \sinh \chi \cos \theta \partial_t - (\cot t \cosh \chi \cos \theta - \sin\theta)\partial_\chi  \nonumber \\ & & + (  coth \chi \cos \theta+\cot t cosech \chi \sin\theta)\partial_\theta, \nonumber \\
 W_2 &=&  - \sinh \chi \sin\theta \partial_t + ( \cot t \cosh \chi \sin\theta + \cos \theta)\partial_\chi \nonumber \\ & &  - ( coth \chi \sin\theta-\cot t cosech \chi \cos \theta )\partial_\theta. \nonumber
\end{eqnarray}
The corresponding basis 1-forms $\{ \bar{e}^a \}$  differ from the right-invariant 1-forms $\{\tilde{e}^a \}$ by an over-all  minus sign: 
\begin{eqnarray}
\bar{e}^0 & =&- \cosh \chi dt - \cos t \sin t \sinh \chi d\chi + \sin^2 t \sinh^2 \chi d\theta,\\
\bar{e}^1 &=& -\sinh \chi \cos \theta dt + (- \cos t \sin t \cosh \chi \cos \theta + \sin^2 t \sin\theta)d\chi \nonumber\\
 &+& \sin^2 t  \sinh \chi( \cosh \chi \cos \theta + \cot t  \sin\theta)d\theta,\\
\bar{e}^2 &=& \sinh \chi \sin\theta dt +( \cos t \sin t \cosh \chi \sin\theta + \sin^2 t \cos \theta)d\chi \nonumber\\
 &-& \sin^2 t  \sinh \chi( \cosh \chi \sin\theta - \cot t  \cos \theta)d\theta,
\end{eqnarray}
and satisfy the following structure equations:
\begin{equation}
d\bar{e}^a =  \epsilon^{a}_{\;\; bc} \bar{e}^b \wedge \bar{e}^c .
\end{equation}
%%%%%%%%%%%%%%%%%%%%%%%%%%%%%%%%%%%%%%%%%%%%%%%%%%%%%%%%%%%%%%%%%%%%%%%

%%%%%%%%%%%%%%%%%%%%%%%%%%%%%%%%%%%%%%%%%%%%%%%%%%%%%%%%%%%%%%%%%%%%%%%%%%


\newpage


 \begin{thebibliography}{99}
 \bibitem{Boulware-Deser} D.Boulware,S.Deser, {\sl Can gravitation have a finite range?}, Phys.Rev.{\bf D6} (1972)  3368
\bibitem{leutwyler} H.Leutwyler, {\sl A 2+1 dimensional model for the quantum theory of gravity}, Nuo.Cim.{\bf 42A}(1966)159
\bibitem{deser-jackiw-tHooft} S.Deser,R.Jackiw,G.'t Hooft, {\sl Three dimensional Einstein gravity: Dynamics of flat space}, Ann.Phys.{\bf 152}(1984) 220; ibid, {\bf 153}(1984) 405
\bibitem{deser-jackiw-templeton2} S.Deser,R.Jackiw,S.Templeton, {\sl Topologically massive gauge theories}, Phys.Rev.Lett. {\bf 48}(1982) 975
\bibitem{deser-jackiw-templeton1} S.Deser,R.Jackiw,S.Templeton, {\sl Topologically massive gauge theories}, Ann.Phys. {\bf 140}(1982) 372,Err. {\sl ibid} {\bf 185}(1988)406
\bibitem{dereli-tucker1}  T.Dereli,R.W.Tucker,{\sl Gravitational interactions in 2+1 dimensions}, Class.Q.Grav.{\bf 5}(1988)951
\bibitem{dereli-deser} T.Dereli,S.Deser, {\sl Fermionic Goldstone-Higgs effect in 2+1 dimensions}, J.Phys.{\bf A11}(1978) L27
%%%%%%%%%%%%%%%%%%%%%%%%%%%%%%%%%%%%%%%%%%%
\bibitem{BTZ1} M.Banados, C.Teitelboim, J.Zanelli, {\sl The black hole in three dimensional space-time},  Phys.Rev.Lett.{\bf 69}(1992)1849
\bibitem{BTZ2} M.Banados, M.Henneaux,  C.Teitelboim, J.Zanelli, {\sl Geometry of the 2+1  black hole },  Phys.Rev.{\bf D48}(1993)1506
%%%%%%%%%%%%%%%%%%%%%%%%%%%%%%%%%%%%%%%%%%%%%%%%%%
\bibitem{brown-henneaux} J.D.Brown, M.Henneaux,{\sl Central charges in the canonical realization of asymptotical symmetries: An example from three dimensional gravity}, Comm.Math.Phys.{\bf 104}(1986) 207 
\bibitem{deser-tekin2}  S.Deser, B.Tekin,  {\sl Energy in topologically massive gravity},  Class.Q.Grav.{\bf 20} (2003)  L259
\bibitem{dereli-obukhov} T.Dereli,Yu.N.Obukhov, {\sl General analysis of self-dual solutions of Einstein-Maxwell-Chern-Simons theory in (1+2) dimensions}, Phys.Rev.{\bf D62}(2000) 024013
\bibitem {carlip et al1}  S.Carlip,S.Deser,A.Waldron,D.K.Wise, {\sl Topologically massive AdS gravity}, Phys.Lett.{\bf B666} (2008) 272 
%%%%%%%%%%%%%%%%%%%%%%%%%%%%%%%%%%%%%%%%%%%%%%%%%%%%%%%%%%%%%%%%%%
\bibitem{Deser1} S.Deser, {\sl Ghost-free, finite 4th order D=3 gravity}, Phys.Rev.Lett. {\bf 103}(2009) 101302
\bibitem{strominger et al} W.Li, W.Song, A.Strominger, {\sl Chiral gravity in three dimensions},JHEP 04 (2008) 082
\bibitem{nakasone-oda} M.Nakasone, I.Oda, {\sl On unitarity of massive gravity in three dimensions}, Prog.Theo.Phys.{\bf 121}(2009) 1389
%%%%%%%%%%%%%%%%%%%%%%%%%%%%%%%%%%%%%%%%%%
\bibitem{bergshoeff-hohm-townsend1} E.A.Bergshoeff, O.Hohm,P.K.Townsend,{\sl Massive gravity in three dimensions}, Phys.Rev.Lett.{\bf 102}(2009)201301
\bibitem{bergshoeff-hohm-townsend2} E.A.Bergshoeff, O.Hohm,P.K.Townsend,{\sl More on massive 3D gravity}, Phys.Rev.{\bf D79}(2009)124040
\bibitem{bergshoeff et al4} H.R.Afshar, E.A.Bergshoeff,W.Merbis,{\sl Extended massive gravity in three dimensions}, JHEP 08 (2014)115  
%%%%%%%%%%%%%%%%%%%%%%%%%%%%%%%%%%%%%%%%%%%%%%%%%%%%%
 \bibitem{bergshoeff et al3} E.A.Bergshoeff, O.Hohm, W.Merbis, A.J.Routh, P.K.Townsend,{\sl Minimal massive 3D gravity}, Class.Q.Grav.{\bf 31}(2014)145008 
 \bibitem{baykal} A.Baykal, {\sl An alternative derivation of the minimal massive 3D gravity}, Clas.Q.Grav.{\bf 32}(2015)025013
 \bibitem{tekin} B.Tekin, {\sl Minimal massive gravity: Conserved charges, excitations, and the chiral gravity limit}, Phys.Rev. {\bf D90} (2014) 081701(R)  
\bibitem{Alishahiha et al} M.Alishahiha, M.M.Qaemaqami, A.Haseh, A.Shirzadi, {\sl On 3D minimal massive gravity}, JHEP 12 (2014) 033
\bibitem{altas-tekin2} E.Altas, B.Tekin,{\sl  Holographically viable extensions of topologically massive and minimal massive gravity}, Phys.Rev.{\bf D93} (2016)025032
\bibitem{arvanitakis} A.S.Arvanitakis,{\sl On solutions of minimal massive 3D gravity}, Class.Q.Grav.{\bf 32}(2015)115010 
\bibitem{altas-tekin1} E.Altas, B.Tekin,{\sl Exact solutions and the consistency of 3D minimal massive gravity}, Phys.Rev.{\bf D92} (2015)025033
\bibitem{cartan} \'{E}.Cartan, {\sl Sur une g\'{e}n\'{e}ralisation de la notion de courbure de Riemann et les espaces $\grave{a}$ torsion}, 
C.R.Acad.Sci.(Paris){\bf 174}(1922)593
\bibitem{trautman} A.Trautman, {\sl Einstein-Cartan theory}, Encyclopedia of Mathematical Physics, edited by J.-P.Fran\c{c}oise,G.L.Naber and S.T.Tsou (Elsevier) {\bf Vol.2}(2006)189 
\bibitem{hehl} M.Lazar,F.W.Hehl,{\sl Cartan's spiral staircase in physics,and in particular,in the gauge theory of dislocations}, Found.Phys.{\bf 490}(2010)1298 
\bibitem{dereli-vercin1}  T.Dereli,A.Ver\c{c}in, {\sl A gauge model of amorphous solids containing defects}, Phil.Mag. {\bf B56}(1987)625
\bibitem{dereli-vercin2} T.Dereli,A.Ver\c{c}in, {\sl A gauge model of amorphous solids containing defects II.Chern-Simons free energy}, Phil.Mag.{\bf B64}(1991)509
\end{thebibliography}
\end{document}